\documentclass[12pt]{article}
\usepackage[dvips]{graphicx}
\usepackage{amsmath,amssymb}
\usepackage{bm}

\title{BSD2}
\author{}

\begin{document}
{\pagestyle{empty}
\rightline{November 2008}
\rightline{~~~~~~~~}
\vskip 1cm
\centerline{\large \bf Regularized Euler product for the zeta function and}
\centerline{\large \bf the Birch and Swinnerton-Dyer and the Beilinson conjecture}
\vskip 1cm
\centerline{{Minoru Fujimoto\footnote{E-mail address: 
             cayce@eos.ocn.ne.jp}} and
            {Kunihiko Uehara\footnote{E-mail address: 
             uehara@tezukayama-u.ac.jp}}}
\vskip 1cm
\centerline{\it ${}^1$Seika Science Research Laboratory,
Seika-cho, Kyoto 619-0237, Japan}
\centerline{\it ${}^2$Department of Physics, Tezukayama University,
Nara 631-8501, Japan}
\vskip 2cm

\centerline{\bf Abstract}
\vskip 0.2in

  We present another expression to regularize the Euler product representation of 
the Riemann zeta function. 
The expression itself is essentially same as the usual Euler product 
that is the infinite product, but we define a new one as the limit of the product 
of some terms derived from the usual Euler product. 
We also refer to the relation between the Bernoulli number and $P(z)$, 
which is an infinite summation of a $z$ power of the inverse primes. 
When we apply the same technique to the  $L$-function associated to an elliptic curve, 
we can evaluate the power of the Taylor expansion for the function even in 
the critical strip, 
which is deeply related to problems known as the Birch and Swinnerton-Dyer 
conjecture and the Beilinson conjecture. 

\vskip 0.4cm\noindent
PACS number(s): 02.30.-f, 02.30.Gp, 05.40.-a

\hfil
\vfill
\newpage}
\setcounter{equation}{0}
\addtocounter{section}{0}
\section{Introduction}
\hspace{\parindent}

  In the situation of successful applications to the physics, 
the zeta function have been used for convenient regularizations 
for divergent calculations in the physics. 
In the previous work\cite{Fujimoto3}, we developed a regularization for 
the Euler product representation by way of the expansion of the logarithm of 
the zeta function but not using M\"obius function 
like the Artin-Hasse exponential\cite{Koblitz}.
This was an effort to search the regularized expression for the Euler product 
of the zeta function in the critical strip $0<\Re z<1$, 
whereas the usual definition of the Riemann zeta function is
\begin{equation}
  \zeta(z)=\sum_{n=1}^\infty\frac{1}{n^z}
          =\prod_{n=1}^\infty\left(1-\frac{1}{{p_n}^z}\right)^{-1}
\label{e101}
\end{equation}
for $\Re z>1$, where the right hand side is the Euler product representation 
and $p_n$ is the $n$-th prime number. 
  In this paper, we continue to deal with the Euler product of the zeta function 
to well-define even in the critical strip $0<\Re z<1$, such as the alternating 
summation in the case of the summation definition:
\begin{equation}
  \hat\zeta(z)=\frac{1}{1-2^{1-z}}\sum_{n=1}^\infty\frac{(-1)^{n-1}}{n^z},
\label{e102}
\end{equation}
where we adopt a hat notation as in $\hat{\zeta}(z)$ for the regularized 
functions, 
which is well defined even in the critical strip $0<\Re z<1$. 
In the same way as the previous paper, we are only interested in the region $\Re z\ge\frac{1}{2}$ 
for the Riemann zeta function, because the functional equation ensures us that 
the regularized nature of the zeta function for the other half plane $\Re z<\frac{1}{2}$. 

  In order to get the regularized expression of the Euler product for the Riemann zeta function, 
we consider another expression defined by the limit of some expressions in section 2. 
In section 3, we argue a relation between the Bernoulli number and 
$\displaystyle{P(z)=\sum_{n=1}^\infty{p_n}^{-z}}$.
When we develop same technique for regularized expression applied to the Euler product 
for the $L$-function associated to an elliptic curve, 
we can get the Taylor expansion at the pole in critical strip 
which is deeply related to the Birch and Swinnerton-Dyer (BSD) conjecture and the Beilinson conjecture. 
So in \S 4 we discuss the BSD and the Beilinson conjecture and concluding remarks.

\vskip 5mm
\section{Another expression for the Euler product representation}
\hspace{\parindent}

  When we think about the Euler's alternating series
\begin{equation}
  \hat{\zeta}(z)=\frac{1}{1-2^{1-z}}\sum_{k=1}^\infty\frac{(-1)^{k-1}}{k^z}
\label{e201}
\end{equation}
for the Riemann zeta function, 
we can evaluate the function for the valuable $z$ even in the critical strip 
$0<\Re z<1$ as well as $1<\Re z$. 
We search for a similar expression for the Euler product of the Riemann 
zeta function in this section. 

  In the same way as the previous work\cite{Fujimoto2}, here we define the $n$-term summation 
or product which converges to the Riemann zeta function in the limit of $n\to\infty$,
\begin{eqnarray}
  \zeta_n^\Sigma(z)&=&\sum_{k=1}^n\frac{1}{k^z}\\
  \zeta_n^\Pi(z)   &=&\prod_{k=1}^n\left(1-\frac{1}{{p_k}^z}\right)^{-1}
\label{e202}
\end{eqnarray}
and
\begin{equation}
  \zeta(z)=\lim_{n\to\infty}\zeta_n^\Sigma(z)
          =\lim_{n\to\infty}\zeta_n^\Pi(z)
\label{e203}
\end{equation}
for $\Re z>1$. 
For the hat notation, we write well-regularized zeta function 
even in the critical strip $0<\Re z<1$ as
\begin{equation}
  \hat\zeta(z)=\lim_{n\rightarrow\infty}\hat{\zeta}_n(z)
              =\lim_{n\rightarrow\infty}\frac{1}{1-2^{1-z}}\xi_n(z),
\label{e204}
\end{equation}
where $\xi_n(z)$ is defined by $\displaystyle{\xi_n(z)=\sum_{k=1}^n\frac{(-1)^{k-1}}{k^z}}$.

  The relation between $\zeta_n^\Sigma(z)$ and $\zeta_n^\Pi(z)$ was also discussed in
the previous work\cite{Fujimoto2} and we evaluate as
\begin{eqnarray}
  \zeta_n^\Sigma(z)&=&\zeta_n^\Pi(z)+\sigma_n(z)\nonumber\\
  \xi_{2n}(z)&=&\zeta_{2n}^\Sigma(z)-2^{1-z}\zeta_n^\Sigma(z),
\label{e205}
\end{eqnarray}
where $\sigma_n(z)$ becomes zero for $n\to\infty$. 
We can easily find an expression of $\xi(z)$ using these two equations above as
\begin{eqnarray}
  \xi(z)\equiv
    \lim_{n\to\infty}\xi_{2n}(z)&=&\lim_{n\to\infty}\{\zeta_{2n}^\Pi(z)-2^{1-z}\zeta_n^\Pi(z)\}\nonumber\\
    &=&\lim_{n\to\infty}\left(
      \prod_{k=1}^{2n}\frac{1}{1-p_k^{-z}}-2^{1-z}\prod_{k=1}^n\frac{1}{1-p_k^{-z}}
    \right)\nonumber\\
    &=&\lim_{n\to\infty}\left\{
    \zeta_n^\Pi(z)\left(\prod_{k=n+1}^{2n}\frac{1}{1-p_k^{-z}}-2^{1-z}\right)\right\},
\label{e206}
\end{eqnarray}
where we have used the fact that $\displaystyle{\lim_{n\to\infty}\sigma_n(z)=0}$ in Eq.(\ref{e205}).

  Each zero is the solution for $\hat{\zeta}(z)=0$ in the region of $\xi(z)=0$ for $|z|>0$ 
because of the last line in Eq.(\ref{e206}).
  Each solution for $\hat{\zeta}_n(z)=0$ is also the solution $\rho$ for $\xi_{2n}(z)=0$ 
for $\zeta_n^\Pi(z)\ne 0$.
So the condition for $\rho$ is 
\begin{equation}
  \prod_{k=n+1}^{2n}(1-p_k^{-\rho})=2^{\rho-1}+o(n^{\rho-1})
\label{e208}
\end{equation}
and for $z\ne\rho$, namely $\xi_{2n}(z)\ne 0$, the relation 
\begin{equation}
  \prod_{k=n+1}^{2n}(1-p_k^{-z})=2^{z-1}+O(n^{z-1})
\label{e209}
\end{equation}
is satisfied, where 
$\displaystyle{\zeta_n^\Pi(z)=\hat{\zeta}_n(z)+\frac{n^{1-z}}{1-z}-\sigma_n(z)}$.\cite{Fujimoto3} 
\begin{eqnarray}
  \lim_{n\to\infty}\sum_{k=n+1}^{2n}\log(1-p_k^{-\rho})&=&(\rho-1)\log 2\\
  \lim_{n\to\infty}\sum_{k=n+1}^{2n}\log(1-p_k^{-z})&=&(z-1)\log 2+O(-\log(z-1))
\label{e210}
\end{eqnarray}

  Under the Riemann hypothesis, we get
\begin{equation}
  \sum_{k=n+1}^{2n}\left(
    \frac{1}{p_k^{\rho}}+\frac{1}{2}\frac{1}{p_k^{2\rho}}
                          +\frac{1}{3}\frac{1}{p_k^{3\rho}}+\cdots\right)=(1-\rho)\log 2
\label{e211}
\end{equation}
and we add this equation and the other given by substituting $1-\rho$ for $\rho$, 
we get finally 
\begin{equation}
  \sum_{k=n+1}^{2n}
    \left\{
      \left(\frac{1}{p_k^{\rho}}+\frac{1}{p_k^{1-\rho}}\right)+
      \frac{1}{2}\left(\frac{1}{p_k^{2\rho}}+\frac{1}{p_k^{1-2\rho}}\right)+\cdots
    \right\}=\log 2.
\label{e212}
\end{equation}

  In order to refer to the condition of the zeros on the critical line for the Riemann hypothesis, 
we introduce the ratio $f_n(z)$ of two products as
\begin{eqnarray}
  f_n(z)&\equiv&\frac{\zeta_{2n}^\Pi(z)}{\zeta_{n}^\Pi(z)}
         =\prod_{r=n+1}^{2n}\frac{1}{1-{p_r}^{-z}}\ll\infty\\
        &\simeq&\frac{\zeta_{2n}^\Sigma(z)}{\zeta_{n}^\Sigma(z)}
         =\frac{\xi_n(z)+2^{1-z}\zeta_n^\Sigma(z)}{\zeta_{n}^\Sigma(z)}
        \simeq 2^{1-z},
\label{e213}
\end{eqnarray}
where we made use of Eq.(\ref{e206}).

  The distinctive features of $f_n(z)$ for the Euler product are 
\begin{enumerate}
  \item We get a finite expression of the Euler product for $|z|<1$, 
        whereas the usual one diverges in the critical strip.
  \item We can decide the value of regularized zeta function itself.
        \begin{equation}
           \hat{\zeta}_n(z)=
             \frac{\zeta_{2n}^\Pi(z)-2^{1-z}\zeta_n^\Pi(z)}
                  {\zeta_{2n}^\Pi(z)-\zeta_n^\Pi(z)}
             \frac{n^{1-z}}{z-1}
        \label{e214}
        \end{equation}
  \item We cat get the expression of $\hat{\zeta}_n(z)$ by way of $f_n(z)$. 
        Thus it can be regarded the condition $f_n(z)=0$ of regularized quantities 
        as the Euler product expression for $\zeta_n(z)$. 
        \begin{equation}
          \hat{\zeta}_n(z)=
            \frac{f_n(z)-2^{1-z}}{f_n(z)-1}\frac{n^{1-z}}{z-1}
        \label{e215}
        \end{equation}
\end{enumerate}

  After all, we conclude that
\begin{eqnarray}
  f_n(z)&\equiv&\frac{\zeta_{2n}^\Pi(z)}{\zeta_{n}^\Pi(z)}=2^{1-z}+O(n^{z-1})\nonumber\\
  \lim_{n\to\infty}f_n(\rho)&=&2^{1-\rho}.
\label{e216}
\end{eqnarray}
Thanks to these expressions, we have become to be able to evaluate a limit of finite series in 
the Euler product compared with the previous work\cite{Fujimoto1}, in which we regularized 
infinite amounts by way of the dipole limitation method.

\vskip 5mm
\section{The Bernoulli number and the regularized Euler product representation}
\hspace{\parindent}

  As is stated in the previous paper,\cite{Fujimoto3}
we can transform the zeta function to followings by taking logarithms of Eq.(\ref{e101}):
\begin{eqnarray}
\log\zeta(z)&=&\sum_{n=1}^\infty\left(
					\frac{1}{{p_n}^z}+\frac{1}{2{p_n}^{2z}}+
					\frac{1}{3{p_n}^{3z}}+\frac{1}{4{p_n}^{4z}}+
					\frac{1}{5{p_n}^{5z}}+\cdots
			\right)
\nonumber
		\\
		&=&	P(z)+\frac{1}{2}P(2z)+
			\frac{1}{3}P(3z)+\frac{1}{4}P(4z)+
			\frac{1}{5}P(5z)+\cdots
\label{e302}
		\\
		&=&	P(z)+R(z)
\nonumber
		,
\end{eqnarray}
where we have used $\displaystyle{P(z)=\sum_{n=1}^\infty\frac{1}{{p_n}^z}}$ in the second line. 
We easily recognize that each term after the second converges in Eq.(\ref{e302}) 
in the region $\Re z>\frac{1}{2}$
and it can be shown that the infinite sums after the forth term converges 
by using the formula $\displaystyle{\sum_{k=2}^\infty\frac{\zeta(k)-1}{k}=1-\gamma}$ 
with the Euler constant $\gamma$.
Thus we express only the first term $P(z)$ as the term for a regularization 
and the expression for $P(z)$ for $\Re z>1$ can be got by adding or subtracting a term of 
$\displaystyle{\sum\frac{1}{\prod p}\log\zeta(\prod p\ z)}$, 
\begin{equation}
  P(z)=\log\zeta(z)
      -\sum_{i=1}^\infty\frac{1}{p_i}\log\zeta(p_iz)
      +\sum_{1\le i<j}^\infty\frac{1}{p_ip_j}\log\zeta(p_ip_jz)
      -\sum_{1\le i<j<k}^\infty\frac{1}{p_ip_jp_k}\log\zeta(p_ip_jp_kz)+-\cdots,
\label{e305}
\end{equation}
where a coefficient of $\displaystyle{\frac{1}{m}P(mz)}$ term 
for $m={p_{i_1}}^{\alpha_1}{p_{i_2}}^{\alpha_2}\cdots{p_{i_n}}^{\alpha_n}$ 
cancels out by using the relation 
$\displaystyle{
  \sum_{i=1}^n(-1)^n
  \left(\begin{array}{@{\,}c@{\,}}n\\i\end{array}\right)=0
}$. 

  A regularization of $P(z)$ in the region of $\frac{1}{2}<\Re z<1$ is also performed 
in the previous work\cite{Fujimoto2,Fujimoto3} where
we have introduced 
\begin{eqnarray}
  \hat{\zeta}(z)=\lim_{n\to\infty}\hat{\zeta}_n(z)
                &=&\lim_{n\to\infty}\left\{\zeta_n^\Sigma(z)-\frac{n^{1-z}}{1-z}\right\},
\label{e308}\\
  \hat{P}(z)=\lim_{n\to\infty}\hat{P}_n(z)
			&=&\lim_{n\to\infty}\left\{P_n(z)-\int_1^n\frac{dt}{(t\log t)^z}\right\}
\label{e310}
\end{eqnarray}
with $\displaystyle{P_n(z)\equiv\sum_{k=1}^n\frac{1}{{p_k}^z}}$, 
and we define the term
\begin{equation}
  R_n(z)\equiv\log\zeta_n^\Sigma(z)-P_n(z),
\label{e311}
\end{equation}
which is finite in the limit of $n\to\infty$. 

The conclusions we have got are 
\begin{eqnarray}
  P_n(z)&=&\log\frac{n^{1-z}}{1-z}+o(n)\\
  \hat{P}(z)&=&\log\frac{1}{1-z}+O(1).
\label{e312}
\end{eqnarray}
For getting an expression $R_n(z)$ in Eq.(\ref{e311}), 
we have evaluated $\displaystyle{P_n(z)=\int_1^n \frac{dt}{{p_t}^z}+O(1)}$. 
After all for $\frac{1}{2}<\Re z<1$, $\hat{P}(z)$ is finite and 
$\hat{\zeta}(z)=-e^{\{\hat{P}(z)+R(z)\}}$ cannot vanish in this strip. 

  We focus on the relation between the Bernoulli number and $P(z)$ from now on.
We put down the relation
\begin{equation}
  \zeta(1-z)=-\frac{1}{z}\lim_{n\to\infty}\sum_{r=1}^n 
  \left(\begin{array}{@{\,}c@{\,}}z\\r\end{array}\right) b_r
  =-\frac{1}{z}\lim_{n\to\infty}b_n(z)
\label{e315}
\end{equation}
for any $z$, 
where this is only true for $z=2m+1$($m$:positive integer) before the regularization though, 
$\displaystyle{
  \left(\begin{array}{@{\,}c@{\,}}z\\r\end{array}\right)
  \equiv\frac{\Gamma(z+1)}{\Gamma(z-r+1)\Gamma(r+1)}
}$ 
and $b_r$'s are the $r$-th Bernoulli number. 
Here we consult the relation of Eq.(22) in the work\cite{Fujimoto2}
\begin{equation}
  \hat{\zeta}(1-z)=\zeta_n^\Sigma(1-z)+H(1-z)\zeta_n^\Sigma(z)+r_n(1-z),
\label{e316}
\end{equation}
where $\displaystyle{H(z)=2\Gamma(1-z)(2\pi)^{z-1}\sin\frac{\pi z}{2}}$, 
and the residue $r_n(z)$ converges to zero in the limit of $n\to\infty$. 
Taking this limit in Eq.(\ref{e316})
\begin{equation}
  \lim_{n\to\infty}\hat{\zeta}_n(1-z)=\lim_{n\to\infty}\frac{-b_n(z)-n^z}{z}
                                     =-\lim_{n\to\infty}\frac{\hat{b}_n(z)}{z},
\label{e318}
\end{equation}
namely we get
\begin{equation}
  b(z)\equiv\lim_{n\to\infty}\hat{b}_n(z)
        =-\lim_{n\to\infty} z\hat{\zeta}_n(1-z)
        =-z\hat{\zeta}(1-z).
\label{e319}
\end{equation}
We can write down the relations using Eqs.(\ref{e302}),(\ref{e312}) and (\ref{e319})
\begin{equation}
  b(1-z)=e^{R(z)},\ R(z)=\log b(1-z)
\label{e320}
\end{equation}
and
\begin{eqnarray}
  b(z)&=&-z\hat{\zeta}(1-z)
        =-z\hat{\zeta}(z)\Gamma(z)2^{1-z}\pi^{-z}\cos\frac{\pi z}{2}\nonumber\\
      &=&\left[\sum_{r=0}^\infty
       \left(\begin{array}{@{\,}c@{\,}}z\\r\end{array}\right) b_r\right]^{DC}\nonumber\\
      &=&\left[1-\frac{z}{2}+\frac{z(z-1)}{2}B_1-\frac{z(z-1)(z-2)(z-3)}{24}B_2\right.
\label{e361}\\
      &&+\frac{z(z-1)(z-2)(z-3)(z-4)(z-5)}{720}B_3\nonumber\\
      &&\left.-\frac{z(z-1)(z-2)(z-3)(z-4)(z-5)(z-6)(z-7)}{40320}B_4+\cdots\right]^{DC},\nonumber
\end{eqnarray}
where $\displaystyle{[\cdots]^{DC}}$ means that a regularization\cite{Fujimoto1} 
is performed inside the parentheses 
and $B_k$'s are also called the Bernoulli numbers whose values are 
$B_1=\frac{1}{6},B_2=\frac{1}{30},B_3=\frac{1}{42},B_4=\frac{1}{30}$ and so on. 
Here we present the formula for the Riemann zeta function by way of $b(z)$ given by 
\begin{equation}
  \hat{\zeta}(z)=-b(1-z)e^{\hat{P}(z)}.
\label{e321}
\end{equation}
Some terms of $n$-th term for $b(z)$ defined in Eq.(\ref{e315}) are
\begin{eqnarray}
  b_0(z)&=&1\nonumber\\
  b_1(z)&=&1\nonumber\\
  b_2(z)&=&1-\frac{1}{2}+\frac{z(z-1)}{12}=\frac{1}{12}(z-3)(z-4)\nonumber\\
  b_4(z)&=&\frac{1}{12}(z-3)(z-4)-\frac{1}{720}z(z-1)(z-2)(z-3)\nonumber\\
        &=&-\frac{1}{720}(z-3)(z-5)(z-6)(z+8)\\
  b_6(z)&=&\frac{1}{30240}(z-3)(z-5)(z-7)(z-8)(z^2+8z+36)\nonumber\\
        &\cdots&\nonumber
\label{e362}
\end{eqnarray}
and a general $2n$-th term of $b(z)$ is
\begin{equation}
  b_{2n}(z)=\frac{b_{2n}}{(2n)!}(z-3)(z-5)(z-7)\cdots\{z-(2n-1)\}(z-2n)g_{2n}(z),
\label{e363}
\end{equation}
where $g_n(z)$'s are given by
\begin{eqnarray}
  g_1(z)&=&-2\nonumber\\
  g_2(z)&=&1\nonumber\\
  g_4(z)&=&z+8\\
  g_6(z)&=&z^2+8z+36\nonumber\\
        &\cdots&.\nonumber
\label{e364}
\end{eqnarray}
As we factorized trivial zeros out from $b_{2n}(z)$, solutions of the limit of the regularized $[g_{2n}(z)]^{DC}=0$, 
namely zeros of $\lim_{n\to\infty}\hat{g}_{2n}(z)=0$ 
should be identical to non-trivial zeros of the Riemann zeta function, 
which means that we could get an Hadamard expression of the Riemann zeta function 
by way of the Bernoulli numbers and same of the $L$-functions. 

  The reason why we can get the relations stated in this section 
without the M\"obius function is deeply related to an expression using by primes 
like Eq.(\ref{e305}).
When we assume the expression
\begin{equation}
  f(x)=\sum_{n=1}^\infty n^\alpha g(n^\beta x)
\end{equation}
is absolute convergent, we get the expression for $g(x)$ by way of $f(x)$ and primes as
\begin{equation}
  g(x)=f(x)-\sum p^x f(p^\beta x)
           +\sum_{p_1<p_2}(p_1p_2)^\alpha f((p_1p_2)^\beta x)
           -\sum_{p_1<p_2<p_3}(p_1p_2p_3)^\alpha f((p_1p_2p_3)^\beta x)+-\cdots,
\end{equation}
where we suppress the residue as we like taking a sufficient large prime $p_M$($p_1<p_2<\cdots<p_M$). 
In this way we have got Eq.(\ref{e305}), where we can get the relation by putting $z=0$ 
\begin{eqnarray}
  P(0)&=&\log\zeta(0)
      -\sum_{i}\frac{1}{p_i}\log\zeta(0)
      +\sum_{i<j}\frac{1}{p_ip_j}\log\zeta(0)
      -\sum_{i<j<k}\frac{1}{p_ip_jp_k}\log\zeta(0)+-\cdots\nonumber\\
      &=&\log\zeta(0)\prod_{k=1}^\infty\left(1-\frac{1}{p_k}\right)\nonumber\\
      &=&\log\left(-\frac{1}{2}\right)\cdot 0=0.
\end{eqnarray}
So this formula is well defined even in $z=0$ which should be compared with $\hat{P}(0)=0$. 
Some other values for $P(z)$ of an integer $z$ are got by Eq.(\ref{e321}) like 
$P(-1)=-\log 2,\ P(-3)=-\log 4$.

\vskip 5mm
\section{The BSD conjecture and the Beilinson conjecture}
\hspace{\parindent}
  The BSD conjecture\cite{Birch1}\cite{Birch2} claims that 
the rank of the abelian group of rational points on an elliptic curve $E$ 
\begin{equation}
  E:Y^2+aY=X^3+bX^2+cX+d
\label{e401}
\end{equation}
is equals to the order of zero of the associated $L$-function 
\begin{equation}
  L(E,z)=\prod_{n=1}^\infty\frac{1}{1-(p_n-N_{p_n}(E))p_n^{-z}+p_n^{1-2z}}
\label{e403}
\end{equation}
at $z=1$ and that 
the equation (\ref{e401}) has infinite rational roots and $L(E,1)=0$ are identical. 
  This conjecture leads to the theorem\cite{Gross}\cite{Kolyvagin} 
that Eq.(\ref{e401}) has finite rational roots in case of $L(E,1)\ne 0$ and that 
Eq.(\ref{e401}) has infinite rational roots in case that 
$L(E,z)=0$ has the order-one zero at $z=1$. 

  When we apply the regularization for the Euler product developed 
in the previous section to the $L$-function associated to 
an elliptic curve, (in which the Riemann hypothesis holds\cite{Fujimoto2},)
the BSD conjecture can be taken into account.

  As is mentioned in the previous work\cite{Fujimoto3}, 
the case of $L(E,1)\ne0$ is followed by the regularization of the Euler product and 
$L'(E,1)\ne0$ in case of $L(E,1)=0$ is satisfied by the generalized Riemann hypothesis. 
Here we express the coefficients numerically of the Taylor expansion of 
the Riemann zeta function at $z=1$, which is deeply related to the Beilinson conjecture. 

  When we expand the Riemann zeta function around $z=1$, 
it gives 
\begin{equation}
  \zeta(z)=\frac{1}{z-1}+\sum_{n=0}^\infty\frac{(-1)^n}{n!}\gamma_n(z-1)^n,
\label{e409}
\end{equation}
where the coefficients $\gamma_n$ are refered to as the Stieltjes constants given by
\begin{equation}
  \gamma_n\equiv\lim_{m\to\infty}
    \left\{\sum_{k=1}^m\frac{(\log k)^n}{k}-\frac{(\log m)^{n+1}}{n+1}\right\}
\label{e410}
\end{equation}
and $\gamma_0$ is known as the Euler constant $\gamma$. 

We introduce the function $Z(n)$ which is the summation the negative integer powers of such zeros by 
\begin{eqnarray}
  Z(n)&\equiv&\sum_k\rho_k^{-n}\nonumber\\
      &=&\sum_{j=1}^\infty
         \left\{\left(\frac{1}{2}+i\lambda_j\right)^{-n}+
                \left(\frac{1}{2}-i\lambda_j\right)^{-n}\right\},
\label{e411}
\end{eqnarray}
where $\rho_k$ is the $k$-th nontrivial zero of $\zeta(z)$ and $\rho_k=\frac{1}{2}\pm i\lambda_j$ 
under the condition of the Riemann hypothesis\cite{Fujimoto1,Fujimoto2} which proof is given in Appendix.

Such summations can be represented by the Stieltjes constants $\gamma_n$ as 
\begin{eqnarray}
  Z(1)&=&\frac{1}{2}(2+\gamma-\log(4\pi))\nonumber\\
  Z(2)&=&1+\gamma^2-\frac{1}{8}\pi^2+2\gamma_1\nonumber\\
  Z(3)&=&1+\gamma^3+3\gamma\gamma_1+\frac{3}{2}\gamma_2-\frac{7}{8}\zeta(3)\nonumber\\
  Z(4)&=&\cdots\nonumber.
\label{e412}
\end{eqnarray}
Using these relations above, we can solve $\gamma_n$ by way of $Z(n+1)$ and $\gamma_k$ where $k<n$ as
\begin{eqnarray}
  \gamma_0&=&2Z(1)-2+\log(4\pi)=\gamma\nonumber\\
  \gamma_1&=&\frac{1}{2}\left(Z(2)-1-\gamma^2-\frac{\pi^2}{8}\right)\nonumber\\
  \gamma_2&=&\frac{2}{3}\left(Z(3)-1-\gamma^3-3\gamma\gamma_1+\frac{7}{8}\zeta(3)\right)\nonumber\\
  \gamma_3&=&\cdots\nonumber
\label{e413}
\end{eqnarray}
  After all every $\gamma_n$ is expressed by $\lambda_j$'s, $\gamma$ and 
the zeta function of odd integers.
This story is easily understood by comparing Eq.(\ref{e409}) with Eq.(\ref{e315}). 
The equation (\ref{e315}) should be read that the Bernoulli numbers are coefficients of 
the Riemann zeta function when it is expanded around $z=1$.
When we look at Eq.(\ref{e361}) replaced $z$ by $z-1$, any coefficient of Eq.(\ref{e409}) 
can be represented by the Bernoulli numbers.

  When the generalized $L$-function is formed by the Euler product expression, 
coefficients of the generalized $L$-function expanded around the pole are given by
the generalized Bernoulli number or the zeros of the generalized Riemann hypothesis. 
This leads us to the goal of the Beilinson conjecture. 

  We know about some examples which show the relation between specials of the zeta function and 
the generalized Bernoulli numbers as
\begin{equation}
  \sum_{(a,b,c)\in S}\frac{\psi(abc)}{(1-\zeta^a)(1-\zeta^b)(1-\zeta^c)}
  =-\sqrt{p}\left(\frac{p+1}{4}B_{1,\psi}+\frac{1}{6}B_{3,\psi}\right),
\label{e329}
\end{equation}
where $S=\{(a,b,c)\in(F_p^\times)^3;ab+bc+ca=0\}$, $\psi$ is the index of quadratic residue, 
$p$ is a prime satisfying $p\equiv3\mod4$ and $\zeta=e^{2\pi i/p}$. 
And some are equalities got by the comparison between the trace formula 
and the Riemann-Roch theorem for the automorphic form,
where generalized Bernoulli numbers often appear. 
We also know that the Dirichlet $L$-function on the integer can be represented 
by the generalized Bernoulli number, which is defined by generating function as 
\begin{equation}
  \sum_{a=1}^f\frac{\chi(a)te^{at}}{e^{ft}-1}=\sum_{n=0}^\infty B_{n,\chi}\frac{t^n}{n!}
\label{e330}
\end{equation}
and the generalized Bernoulli number $B_{n,\chi}$ itself is written by way of 
the Bernoulli polynomial as
\begin{equation}
  B_{n,\chi}=f^{n-1}\sum_{a=1}^f \chi(a)B_n(a/f).
\label{e331}
\end{equation}
Another example of the $L$-function $L(E,z)$ associated with an elliptic curve is also written by 
\begin{equation}
  \frac{L(E,z)}{(z-1)^r}\Big\|_{z=1}=\frac{|\amalg\hspace{-7pt}\amalg(E)|\cdot Reg(E)\cdot \prod_{v:bad}Tam_v}
                                     {|E(Q)_{tor}|^2\cdot \Omega},
\label{e332}
\end{equation}
where $Reg(E)$ is the height pairing, $Tam_v$ is a local Tamagawa number 
and $\Omega$ is the period of $E$.

\vskip 5mm
\newpage
\renewcommand{\theequation}{\Alph{section}\arabic{equation}}
\setcounter{section}{1}
\setcounter{equation}{0}
\section*{Appendix}

\hspace{\parindent}
  Here we present a brief proof of the Riemann hypothesis for the Riemann zeta function. 
As we cited above, the true mechanism of the proof can be seen in previous work\cite{Fujimoto2}, 
we take the essence of the proof by using the functional equation. 
The functional equation is given by
\begin{equation}
  \hat{\zeta}(z)=H(z)\hat{\zeta}(1-z),
\label{ea01}
\end{equation}
where $\displaystyle{H(z)=2\Gamma(1-z)(2\pi)^{z-1}\sin\frac{\pi z}{2}}$.
We easily get the relation
\begin{equation}
  \frac{H(z+2)}{H(z)}=-\frac{4\pi^2}{z(z+1)}
\label{ea02}
\end{equation}
and we can transform it as
\begin{eqnarray}
  z(z+1)&=&-4\pi^2\frac{H(z)}{H(z+2)}
          =-\frac{4\pi^2}{H(z+2)}\frac{\hat{\zeta}(z)}{\hat{\zeta}(1-z)}\nonumber\\
        &=&-\frac{4\pi^2}{H(z+2)}\lim_{n\to\infty}
          \frac{\displaystyle{\frac{\zeta_n(z)}{\sqrt{n}}}-\frac{n^{1/2-z}}{1-z}+o(\sqrt{n})}
               {\displaystyle{\frac{\zeta_n(1-z)}{\sqrt{n}}}-\frac{n^{z-1/2}}{z}+o(\sqrt{n})}\\
        &=&\frac{4\pi^2}{H(z+2)}\frac{z}{z-1}\lim_{n\to\infty}n^{1-2z},\nonumber
\label{ea03}
\end{eqnarray}
where we used the relation derived from Eqs.(7) and (8) in the work\cite{Fujimoto2} such as
\begin{equation}
  \zeta_n(z)=\hat{\zeta}(z)+\frac{n^{1-z}}{1-z}+o(n).
\label{ea04}
\end{equation}

Finally we get the relation 
\begin{equation}
  \frac{(z^2-1)H(z+2)}{4\pi^2z}=\lim_{n\to\infty}n^{1-2z},
\label{ea05}
\end{equation}
where $\Re(1-2z)$ must be equal to zero, because the left hand side is finite for $z\ne\pm 1$.

\vskip 5mm

\end{document}